\documentclass[conference,letterpaper]{IEEEtran}
\usepackage{cite}
\usepackage[utf8]{inputenc}
\usepackage[T1]{fontenc}
\usepackage[english]{babel}
\usepackage{graphicx}
\usepackage{booktabs}
\usepackage[most]{tcolorbox}
\usepackage{enumitem}
\usepackage{microtype}
\usepackage{balance}
\usepackage{hyperref}
\usepackage{tikz}
\usepackage{pgfplots}
\usepackage{stfloats}
\pgfplotsset{compat=1.18}
\hypersetup{colorlinks=true,linkcolor=black,citecolor=black,urlcolor=black}
\usepackage{enumitem}
\setlist[itemize]{leftmargin=*}
\setlist[enumerate]{leftmargin=*}

\title{%Cross-Chiplet Side-Channel Attacks in 2.5/3D Integrated Systems\\or\\
Spying Across Chiplets: Side-Channel Attacks in 2.5/3D Integrated Systems}
\author{
Giorgio Di Natale\IEEEauthorrefmark{1}, 
Christelle Rabache\IEEEauthorrefmark{1}, 
Pierre-Louis Hellier\IEEEauthorrefmark{1}, Florence Podevin\IEEEauthorrefmark{1}, 
Sylvain Bourdel\IEEEauthorrefmark{1}, \\
Romain Siragusa\IEEEauthorrefmark{2}, 
Paolo Maistri\IEEEauthorrefmark{1}  \\
\IEEEauthorblockA{
\IEEEauthorrefmark{1}
Univ. Grenoble Alpes, CNRS, Grenoble INP\textsuperscript{1}, TIMA, 38000 Grenoble, France\\ 
\IEEEauthorrefmark{2}
Univ. Grenoble Alpes, Grenoble INP\textsuperscript{1}, LCIS, 26362 Valence, France\\ 
}
}

\begin{document}
\maketitle

\begin{abstract}
Advanced packaging and chiplet-based integration are increasingly adopted to build complex heterogeneous systems beyond the limits of monolithic scaling. While these architectures offer major benefits in terms of modularity, yield, and performance, they also introduce new physical attack surfaces. In this paper, we show that side-channel attacks can be mounted across chiplets within the same package or stack. Our key idea is that a communication-oriented chiplet, originally intended to interact with the external environment through an antenna, an RFID-like element, or another contactless coupling structure, can be repurposed as an internal observation platform. We formalize this threat through a realistic adversary model, describe the corresponding attack principle, and experimentally assess its feasibility. The obtained results demonstrate that signals captured through such a communication-oriented interface can reveal information correlated with the activity of a neighboring victim chiplet. 
\end{abstract}

% ─────────────────────────────────────────────────────────────
%  SECTION 1 – INTRODUCTION
% ─────────────────────────────────────────────────────────────

\section{Introduction}
\label{sec:introduction}

The continuous slowdown of traditional monolithic scaling has pushed the semiconductor industry toward advanced packaging and heterogeneous integration. Among the most promising solutions, chiplet-based architectures enable complex systems to be assembled from multiple smaller dies, potentially designed in different technologies and optimized for distinct functions such as computation, memory, I/O, or radio-frequency (RF) communication. This modular approach improves yield, reduces development cost, and offers greater design flexibility than large monolithic systems, while also enabling new forms of 2.5D and 3D integration \cite{arm_wireless3d_2019,toshi_2022}.

However, the same features that make chiplets attractive also introduce new security challenges. Splitting a system across multiple dies increases the number of physical interfaces, enlarges the attack surface, and complicates the trust assumptions across the supply chain. Recent studies have highlighted security concerns at several levels of heterogeneous integration, including threats affecting individual chiplets, interposer-level interconnects, and the overall system-in-package. In parallel, prior work has shown that advanced integration technologies can expose new physical leakage paths, such as thermal side channels in 3D integrated circuits and contactless probing opportunities in chiplet-based systems \cite{suzano2024chipletsecurity,knechtel_3d_2017,meyer_chiplet_probing_2024}.

Beyond the chiplet domain, the literature has already demonstrated that side-channel attacks do not necessarily require direct physical access to the victim chip. In shared FPGA platforms, internal sensors implemented in programmable logic have been used to recover secret information from a remote victim implemented elsewhere on the same device. At the board level, related works have further shown that one chip can be used to observe the activity of another chip through shared physical effects, even in the absence of explicit signal-level communication between the attacker and the victim \cite{schellenberg_inside_job_2021,schellenberg_board_2018}. These results suggest that physical isolation at the logical level does not imply side-channel isolation at the system level.

In this paper, we extend this line of reasoning to chiplet-based systems and ask the following question: \emph{can a chiplet be used to perform a side-channel attack against another chiplet within the same package or stack?} This question is especially relevant in emerging architectures where wireless or RF-enabled chiplets are considered for intra-package communication. Recent studies on wireless inter-chiplet communication explore the integration of antennas and transceivers within multi-chiplet systems to reduce wiring constraints and improve communication flexibility \cite{medina_wireless_2023,irabor_wireless_2025}. While such components are primarily introduced for legitimate communication purposes, they may also offer an attacker a privileged physical observation point in very close proximity to the target chiplet.

Our key idea is that an untrusted or malicious RF-capable chiplet can be leveraged not only as a communication component, but also as an electromagnetic sensing interface to observe the activity of a neighboring chiplet. This creates a new attack scenario in which the adversary does not need external probes or direct access to the victim die. Instead, the attack is mounted from within the package itself, using the natural proximity and heterogeneous functionality enabled by chiplet integration.

The main contribution of this work is to demonstrate the feasibility of cross-chiplet side-channel attacks. More specifically, we define a realistic threat model for heterogeneous multi-chiplet systems, propose an attack methodology based on an RF-enabled observer chiplet, and experimentally evaluate the ability of such a chiplet to extract information from a neighboring victim chiplet. We further analyze the conditions under which the attack succeeds, discuss its practical limitations, and outline possible countermeasures at the architectural and packaging levels.

The remainder of this paper is organized as follows. Section~\ref{sec:sota} reviews the relevant literature on chiplet security, side-channel attacks in shared hardware environments, and wireless inter-chiplet communication. Section~\ref{sec:method} presents the threat model and the proposed attack methodology. Section~\ref{sec:results} describes the experimental setup and reports the obtained results. Finally, Section~\ref{sec:limitations} discusses limitations while Section~\ref{sec:conclusion} concludes the paper highlighting future research directions.

\section{Background and Related Work}
\label{sec:sota}

\subsection{Remote and Internal Side-Channel Attacks}

Side-channel analysis has traditionally been associated with direct physical access to the victim device. However, this assumption has been progressively challenged by a series of works showing that information leakage can be observed remotely through shared physical resources. In particular, Zhao and Suh demonstrated that power side-channel attacks can be mounted remotely in shared FPGA platforms, showing that an attacker can recover exploitable leakage without direct access to the victim implementation \cite{zhao2018fpga}. In a closely related direction, Schellenberg \emph{et al.} showed that remote power analysis can also be performed from one FPGA region to another within the same device, despite logical isolation between attacker and victim \cite{schellenberg2021insidejob}.

These results were later extended beyond the single-chip setting. At board level, Schellenberg \emph{et al.} demonstrated that one chip can be used to sense the activity of another chip on the same board, exploiting shared physical effects and the common power delivery environment \cite{schellenberg2018board}. Taken together, these works establish an important lesson: logical isolation does not imply side-channel isolation. Even when two components do not explicitly exchange sensitive data, their physical proximity and shared infrastructure may still create exploitable leakage paths.

For the present work, these studies provide the conceptual foundation. They show that remote and internal side-channel attacks are realistic whenever the attacker can place a sensing structure sufficiently close to the victim and exploit a coupling medium such as the power distribution network or the electromagnetic field. Our work extends this reasoning to chiplet-based systems, where physical proximity is even stronger and where heterogeneous functionality may naturally place an RF-capable chiplet next to a security-sensitive computing chiplet.

\subsection{Security of \texorpdfstring{$2.5$D/3D}{2.5D/3D} Integration and Chiplet-Based Systems}

The transition from monolithic integrated circuits to heterogeneous integration, including $2.5$D interposer-based systems, stacked 3D ICs, and chiplet-based architectures, has introduced new security challenges. Early studies on 3D integration already pointed out that advanced packaging creates new attack surfaces and modifies the threat landscape compared to conventional 2D ICs \cite{dofe2016emerging3d}. Subsequent works further analyzed how vertical integration affects hardware trust, including threats related to hardware Trojans, split trust across the supply chain, and novel physical attack opportunities \cite{knechtel2017thermal3d,rao2022securityvulnerabilities,vashistha2022toshi}.

More recently, security concerns have been revisited specifically in the context of chiplet-based systems. The distributed nature of these systems complicates trust management, broadens the set of potentially untrusted actors, and exposes new interfaces that may be tampered with, monitored, or abused \cite{suzano2024chipletsecurity}. In addition to generic architectural and supply-chain threats, recent studies have emphasized communication-based and physical threats in advanced packages, including covert channels between integrated dies and the vulnerability of chiplet-based platforms to contactless probing techniques \cite{miketic2023covert,deric2024probing}.

Security issues also arise during test and debug. In this regard, DFT-related protection mechanisms have started to receive attention in the chiplet domain, for example through encryption and integrity-checking schemes for secure test access and trustworthy integration \cite{suzano2025dftsecurity}. This trend is important because it confirms that security in chiplet systems is no longer viewed only as a high-level architectural concern, but also as a practical implementation issue affecting interfaces, test infrastructures, and post-silicon operations.

Although these works clearly identify side-channel and physical attacks as relevant threats for advanced integration, they mainly remain at the level of threat analysis, architectural security, probing vulnerability, or defensive mechanisms. They do not explicitly demonstrate a side-channel attack in which one chiplet acts as the observation platform for spying on another chiplet in the same package. This gap motivates the present study.

\subsection{Communication-Oriented Chiplets as a Dual-Use Enabler} \label{subsec:Communication-Oriented_Chiplets}

Recent research has explored wireless communication as a promising complement or alternative to wired interconnects in advanced multi-chiplet systems. In particular, several works have investigated in-package wireless communication to reduce wiring constraints, improve flexibility, and enable efficient communication across heterogeneous chiplets \cite{medina_wireless_2023,irabor_wireless_2025}. These studies show that RF-enabled structures are becoming a realistic option in emerging chiplet-based architectures.

More broadly, heterogeneous chiplet-based systems may naturally integrate communication-oriented dies or interfaces that rely on electromagnetic or contactless interaction with their environment. These may include antennas, RFID-like elements, inductive or capacitive coupling structures, or more generally RF front-ends intended for communication, identification, synchronization, power transfer, or external control. While such components are introduced for legitimate system-level functions, they also constitute physical interfaces placed in very close proximity to other chiplets within the same package or stack.

From a security perspective, this observation is important because such interfaces may provide more than their intended communication function. A chiplet embedding an antenna or another contactless coupling structure may also act as a sensing element, intentionally or unintentionally capturing information correlated with the activity of a neighboring die. In other words, a communication-oriented chiplet may become a privileged observation point for physical leakage generated elsewhere in the system.

This dual-use nature has not been sufficiently investigated in the literature. Existing studies on wireless or RF-assisted chiplet systems mainly emphasize performance, flexibility, and integration benefits. However, the same hardware features that make these solutions attractive from a communication standpoint may also create a new attack surface for side-channel analysis.

In this work, we build precisely on this observation. Rather than restricting the discussion to dedicated intra-chiplet wireless links, we consider a broader and more realistic scenario in which a legitimate communication chiplet, originally intended to communicate with the external world through an antenna, an RFID-like interface, or another contactless coupling structure, is repurposed by an attacker as an observation platform. The key idea is that such a chiplet can be abused to sense leakage from an adjacent victim chiplet, thereby enabling side-channel attacks from within the package itself.

\subsection{Positioning of This Work}

The literature therefore provides three essential ingredients. First, remote side-channel attacks are already known to be feasible in shared FPGA platforms and across chips on the same board \cite{zhao2018fpga,schellenberg_inside_job_2021,schellenberg_board_2018}. Second, chiplet-based and 3D-integrated systems are recognized as introducing new physical and architectural security risks \cite{suzano2024chipletsecurity,meyer_chiplet_probing_2024,toshi_2022,suzano2025dftsecurity}.
Third, heterogeneous chiplet-based systems may naturally integrate communication-oriented dies or interfaces, including RF front-ends, antennas, RFID-like elements, or other contactless coupling structures intended to exchange data with the external environment. While such components are primarily introduced for legitimate communication, control, identification, or low-cost connectivity, they also create an additional physical interface embedded in close proximity to the rest of the system. In some architectures, wireless or contactless communication may even be extended to intra-package links, further reinforcing the presence of RF-enabled structures within advanced multi-chiplet systems \cite{medina_wireless_2023,irabor_wireless_2025}.

What remains missing is the connection between these three lines of work. Existing studies have shown that side-channel observation can be performed remotely, and that chiplet-based systems expose new security-relevant physical interfaces. However, to the best of our knowledge, no prior work has experimentally investigated whether a communication-oriented chiplet, originally intended to interact with the external world, can be repurposed as an observation platform to spy on the activity of a neighboring chiplet within the same package or stack.

This paper fills that gap. Rather than assuming that the attacker necessarily relies on a dedicated intra-chiplet wireless link, we consider a broader and more realistic scenario in which an RF-capable or contactless communication chiplet is legitimately integrated in the system and then abused to sense information leakage from an adjacent victim chiplet. In this sense, the communication interface becomes a dual-use structure: benign from the system designer's perspective, yet exploitable as a side-channel sensor from the attacker's perspective.
\section{Threat Model and Attack Methodology}
\label{sec:method}

\subsection{System Model and Assumptions}
\label{subsec:system_model}

\begin{figure}[t]
    \centering
    \includegraphics[width=\columnwidth]{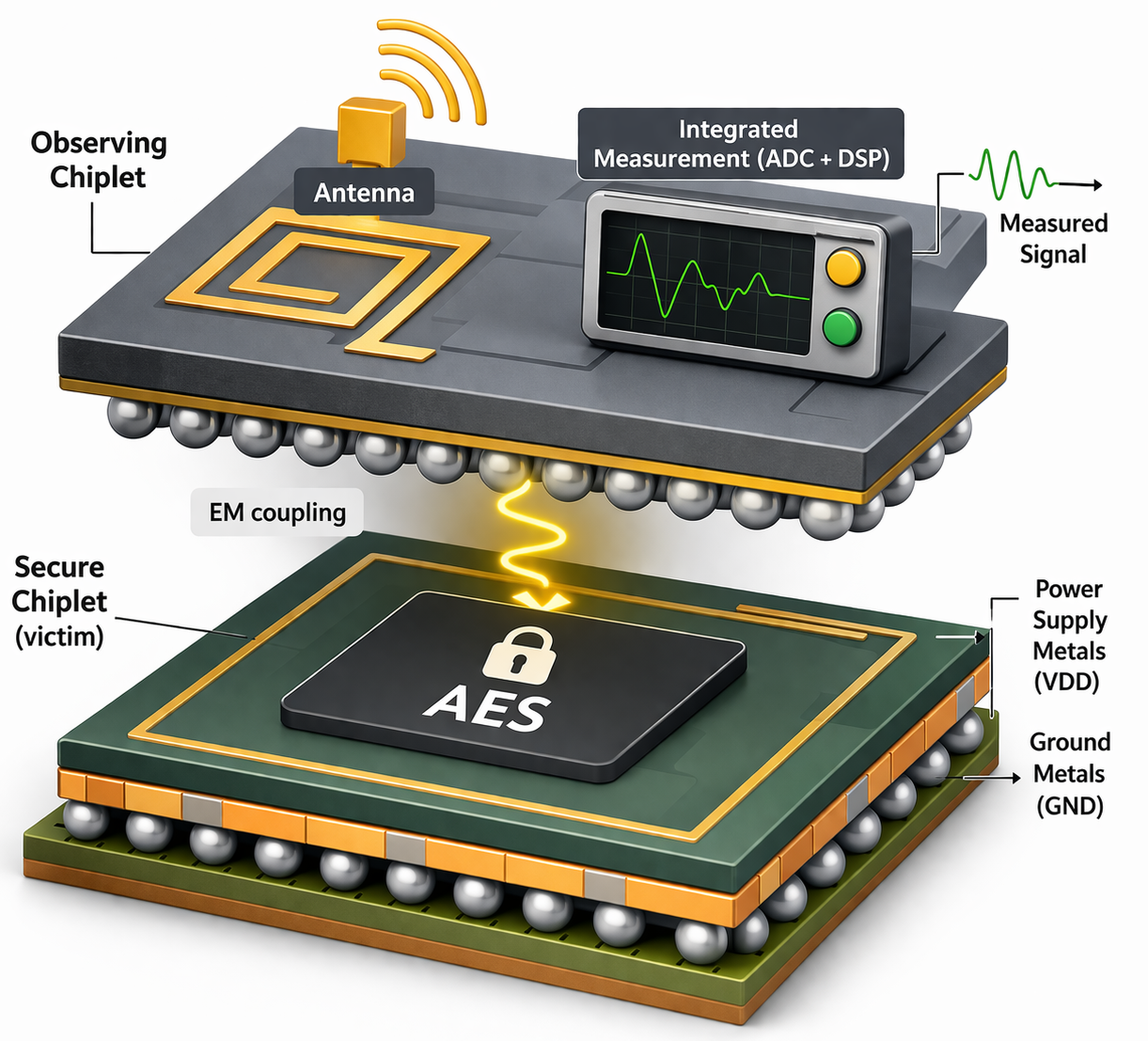}
    \caption{Illustration of the proposed attack scenario.}
    \label{fig:system}
\end{figure}

We consider a heterogeneous multi-chiplet system in which several chiplets are integrated within the same package or stack. The target of the attack is a victim chiplet executing a security-sensitive workload, such as a cryptographic operation or any computation whose internal activity depends on secret data. In close physical proximity to this victim, the system also integrates a communication-oriented chiplet or interface. This second element may be intended for external communication, identification, synchronization, telemetry, or low-power interaction with the environment, and may rely on an antenna, an RFID-like element, or another contactless coupling structure (see Fig. \ref{fig:system}).

The communication-oriented component is assumed to be a legitimate part of the system architecture. Its presence is therefore not suspicious by itself. Importantly, our attack does not require direct electrical access to the victim chiplet, nor invasive probing of its internal nodes. Instead, the attack leverages the fact that both chiplets share the same package or stack and are placed at very short distance from each other, which creates favorable conditions for electromagnetic or contactless coupling.

We further assume that the observing chiplet is able not only to capture analog variations correlated with the victim activity, but also to convert them into a form that can be exploited by the attacker. In particular, the communication-oriented chiplet may embed, or have access to, an acquisition chain including an analog front-end and an analog-to-digital converter (ADC), allowing the observed EM or contactless signal to be transformed into digital waveforms. These measurements may then be extracted through an accessible interface or processed locally, for example by dedicated digital logic or by an embedded processor available within the chiplet.

Our study does not assume a specific packaging technology. The considered scenario is intentionally generic and may apply to different forms of advanced integration, including stacked 3D systems and closely integrated multi-chiplet packages, as long as a communication-capable structure is located sufficiently close to the victim chiplet to capture information correlated with its activity.

\subsection{Adversary Model}
\label{subsec:adversary_model}

We assume an adversary who is able to control, influence, or repurpose the communication-oriented chiplet. This control may arise in different ways. In a first scenario, the communication chiplet itself is untrusted, for instance because it is obtained from a third party or integrated from a less trusted supply-chain source. In a second scenario, the chiplet is benign by design, but its functionality is later misused through malicious firmware, configuration abuse, or unauthorized access to its communication front-end.

The adversary does not have direct access to the victim chiplet's internal signals, secrets, or layout. The attacker is also not assumed to place any external probe above the victim die. Instead, the attacker observes physical effects from within the package by exploiting the legitimate communication structure as an internal sensing interface. In this sense, the attack is mounted from inside the integrated system rather than from the outside.

We further assume that the victim executes the targeted operation multiple times, which is a standard assumption in side-channel analysis. The attacker may know the public inputs associated with these executions, and may use externally observable events for coarse synchronization. However, the attacker does not require any privileged debug access to the victim chiplet.

\subsection{Attack Principle}
\label{subsec:attack_principle}

The core idea of the attack is to repurpose a communication-oriented chiplet as a side-channel sensor. Although originally designed for communication with the external environment, the chiplet's antenna, RFID-like structure, or contactless coupling interface may also capture analog variations correlated with the activity of a neighboring victim chiplet. These variations may originate from electromagnetic emissions, near-field coupling, substrate coupling, package-level propagation, or a combination of multiple physical effects.

When the victim chiplet performs secret-dependent operations, its internal switching activity generates physical leakage. Because of the very short distance between chiplets inside the same package or stack, part of this leakage may be observed by the neighboring communication structure. The captured analog signal is then assumed to be digitized through the available acquisition chain and either exported or processed locally, enabling standard side-channel post-processing and analysis.

The originality of the considered attack lies in the observation point. In conventional side-channel attacks, the attacker places an external probe near the target or exploits a shared infrastructure such as the power distribution network. In our scenario, the sensing element is already embedded inside the system as a legitimate functional block. The communication chiplet therefore acts as a dual-use structure: it fulfills its intended role during normal operation, yet it can also be abused as an internal observation platform.

The goal of the attack is not merely to detect the presence of activity in the victim chiplet, but to determine whether the signals acquired through the communication-oriented structure contain exploitable information about the victim computation. In the following section, we describe the experimental setup used to evaluate this hypothesis and to assess the practical feasibility of cross-chiplet side-channel observation.
\section{Experimental Validation and Results}
\label{sec:results}

This section presents the experimental validation flow used to assess the feasibility of the proposed cross-chiplet side-channel attack. Our objective is to progressively move from a conventional side-channel setting to the targeted chiplet-based scenario, while preserving the same secret-dependent computation and the same attack principle. This provides a clear baseline and allows us to quantify how much information remains exploitable after the propagation and acquisition effects introduced by the second chiplet.

The validation flow is organized in three steps. First, we establish a reference case by performing a side-channel attack from an external observation point, using the supply current of the circuit during the execution of the target operation. Second, these current traces are used as excitation sources in a 3D electrical simulation in order to model the electromagnetic effect induced on a neighboring chiplet and captured through capacitive or inductive coupling. Third, the signal retrieved on the second chiplet is analyzed, %under different acquisition conditions, including multiple temporal resolutions and ADC precisions. Finally, 
and the same attack methodology is applied to the coupled and digitized signal in order to determine whether the secret key can still be recovered. %In this section, we first focus on the reference external attack, which serves as the baseline for the rest of the study.

\subsection{Baseline Validation Through External Current Observation}
\label{subsec:baseline_external_attack}

As a first step, we consider a conventional side-channel setting in which the attacker observes the current consumption of the target circuit directly from the outside. This experiment has two purposes. First, it validates that the considered computation exhibits sufficiently strong data-dependent leakage to support key recovery. Second, it provides the reference traces that will later be injected into the 3D electrical simulation used to emulate the cross-chiplet observation scenario.

The target circuit is implemented in a \(130\,\mathrm{nm}\) technology and consists of a simple cryptographic datapath. An 8-bit input datum \(x\) is XORed with an 8-bit secret key \(k\), and the resulting intermediate value is applied to an AES S-box. The S-box output is connected to a \(5\times 10^{-15}\,\mathrm{F}\) capacitor in order to model the load seen by the circuit. All input values from \texttt{0x00} to \texttt{0xFF} are applied to the circuit, and transient simulations are performed in SPICE. For each input value, the supply current \(I(V_{DD})\) is recorded and used as the side-channel trace.

To evaluate whether the current traces reveal the secret key, we apply a Differential Power Analysis (DPA) procedure based on a prediction of the sensitive intermediate value. For each key hypothesis \(k^\star \in \{0,\ldots,255\}\) and each input \(x\), we compute the hypothetical internal value
\[
v(x,k^\star) = SBox(x \oplus k^\star).
\]
The leakage model is derived from the Hamming weight of this value,
\[
L(x,k^\star)=HW(v(x,k^\star)),
\]
which reflects the number of bits set to logic `1' at the S-box output. Since the S-box output is an 8-bit value, this model captures the aggregate contribution of the eight output bits to the switching activity and therefore provides a natural classifier for the measured traces.

The basic principle of DPA is to compare the measured traces with the leakage predicted for each key hypothesis and to identify the hypothesis that yields the strongest statistical dependency. In practice, for each candidate key, the traces are partitioned or evaluated according to the predicted Hamming-weight classes, and a statistical distinguisher is computed over time. The correct key is expected to produce a significantly stronger peak than the incorrect hypotheses because its predicted intermediate values are consistent with the actual internal computation performed by the circuit.

\begin{figure}[t]
    \centering
    \includegraphics[width=\columnwidth]{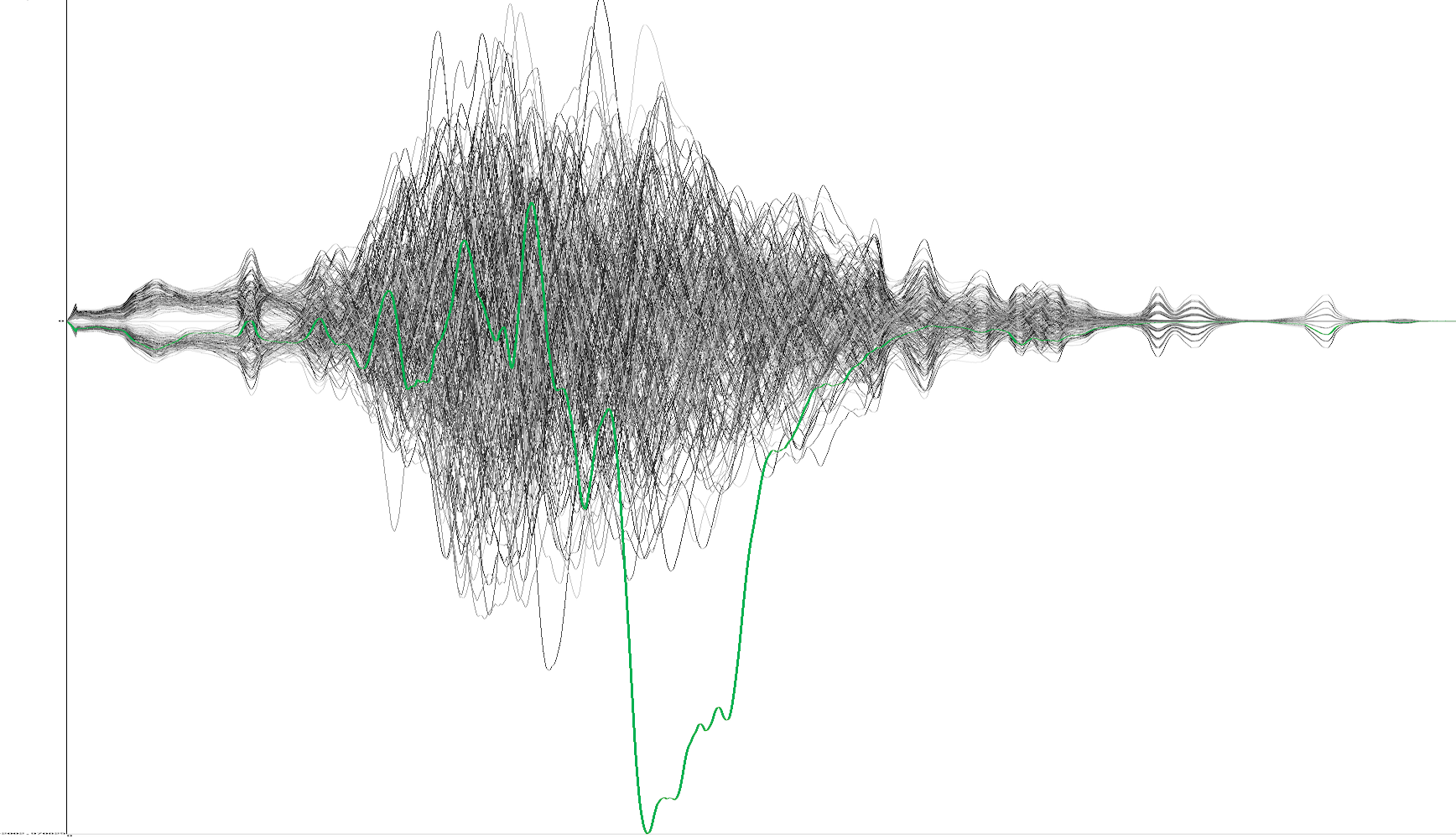}
    \caption{Baseline DPA result obtained from the externally observed supply current \(I(V_{DD})\). The correct key hypothesis exhibits a clear dominant peak, showing that the secret can be recovered from the reference traces.}
    \label{fig:dpa_baseline}
\end{figure}

In our experiments (performed with the tool described in \cite{DPAGio}), this baseline attack is highly successful. Using only the 256 traces corresponding to the 256 possible 8-bit input values, the correct secret key is clearly identified. Figure~\ref{fig:dpa_baseline} illustrates a representative result, in which the distinguisher associated with the correct key hypothesis exhibits a dominant peak, while the wrong hypotheses remain significantly lower. %This first result confirms that the selected circuit and leakage model are suitable for side-channel analysis and provides a strong reference point for the cross-chiplet experiments presented in the next subsections.

\subsection{Electrical Simulation of Cross-Chiplet Electromagnetic Coupling}

To prove that the scenario described in \ref{subsec:system_model} is feasible, an electromagnetic (EM) simulation is performed showing that the observing chiplet can exploit a leakage signal from the victim chiplet (see Fig.~\ref{fig:system}).
The methodology is two-fold. First, an EM model of the system is computed, using the Finite Element Method (FEM) EM solver of RFPro software, provided by the Cadence Suite. Second, a transient simulation is performed to evaluate the leakage signal captured by the observing chiplet.

\begin{figure}[t!]
    \centering
    \includegraphics[width=.5\columnwidth]{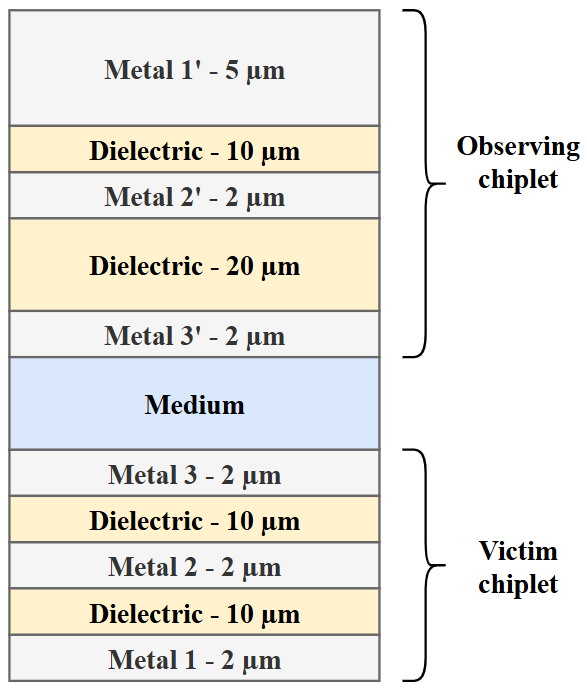}
    \caption{Custom material stack for electromagnetic simulations}
    \label{fig:stack_custom}
\end{figure}

\begin{figure}[t!]
    \centering
    \includegraphics[width=\columnwidth]{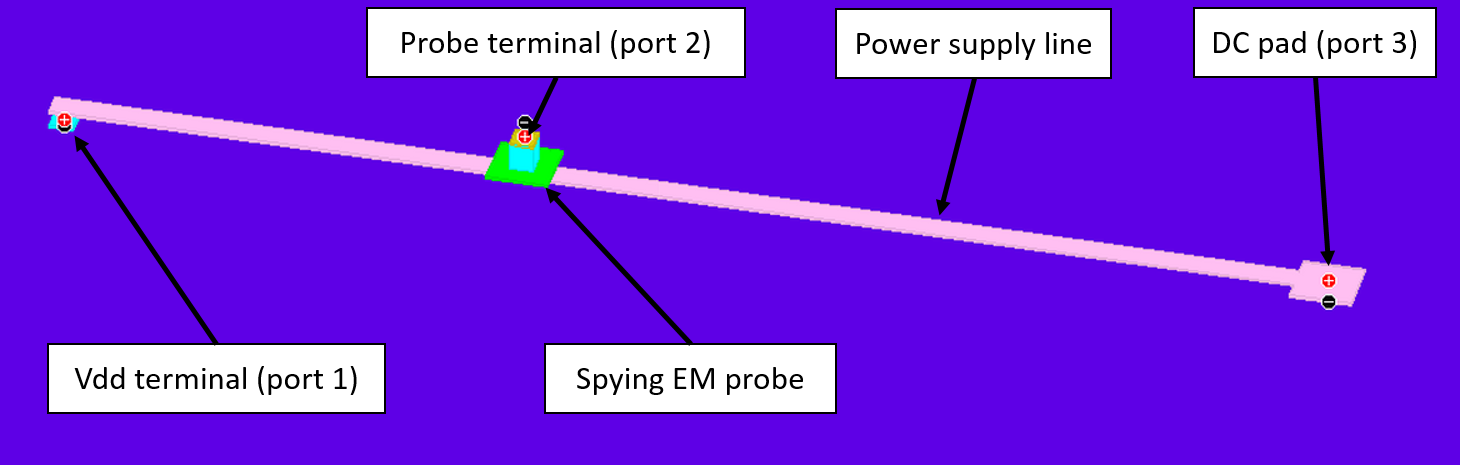}
    \caption{3D view of the system showing the victim chip ground plane (dark blue), power supply line (pink) and the capacitive spying probe (green) which are separated by a non-visible intermediate medium.}
    \label{fig:Capacitive_probe_3Dview}
\end{figure}

\begin{figure}[t!]
    \centering
    \includegraphics[width=\columnwidth]{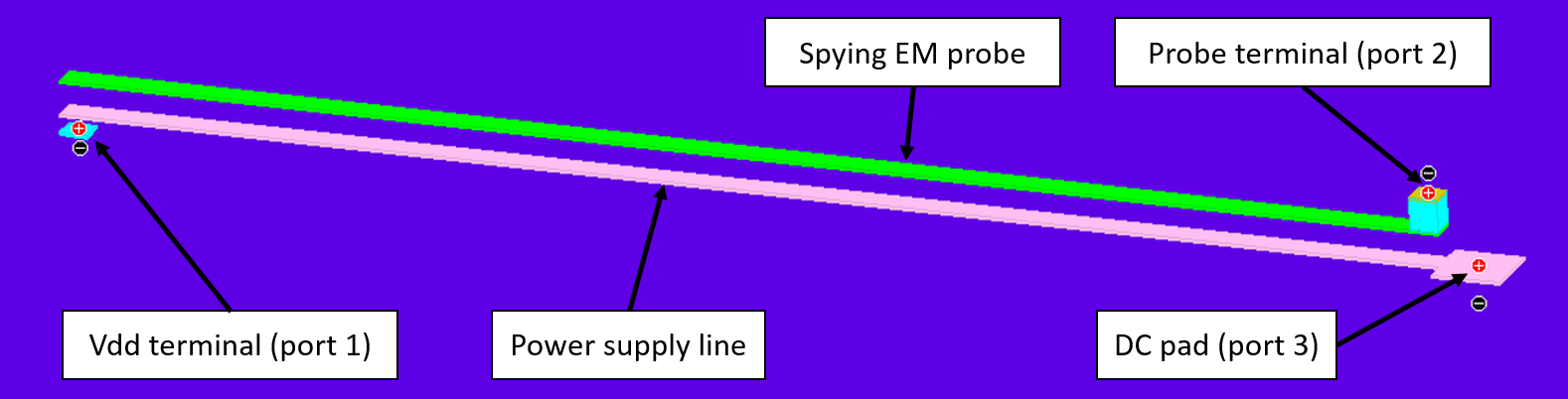}
    \caption{3D view of the system showing the victim chip ground plane (dark blue), power supply line (pink) and the inductive+capacitive spying probe (green) which are separated by a non-visible intermediate medium .}
    \label{fig:Inductive_probe_3Dview}
\end{figure}

A custom stack of materials is described in Fig.~\ref{fig:stack_custom}. Metals 1 to 3 are used to describe the power supply routing of the victim chip. As illustrated in Fig.~\ref{fig:Capacitive_probe_3Dview} and Fig.~\ref{fig:Inductive_probe_3Dview}, the ground plane is in metal 1, the power supply path is in metal 3, presenting a width of 20~µm and a length of 1~mm, and the Vdd terminal of the digital circuit (port~1) is assumed to be in metal 2. It is connected to one end of the power supply line through a via, while a DC pad is connected to the other end (port~3) as in real-life.
Metals 1', 2' and 3' are used to describe the observing chiplet, which is assumed to be positioned up side down compared to the victim chip. Its ground plane is in metal 1', the spying probe is in metal 3', and the measurement circuit input is assumed to be in metal 2' (port~2).

Different hypotheses are explored for the medium between the two chiplets: thin dielectric material, or air gaps of various thicknesses.
Also, two types of spying probe are being considered.
The first type of probe consists of a 50x50~$\mu \text{m}^2$ metal plate which creates a capacitive coupling between the power supply rail and the probe. A 3D view of the system including the capacitive probe is shown in Fig.~\ref{fig:Capacitive_probe_3Dview}. 
The second type of probe consists of a 1-mm long metal line that primarily contributes to an inductive coupling between the power supply rail and the probe but also present capacitive effect as in broadside coupled transmission lines. This use case will be referred to as inductive+capacitive coupling. A 3D view of the system including the inductive+capacitive probe is shown in Fig.~\ref{fig:Inductive_probe_3Dview}, where the probe is aligned with the Vdd line.
As aforementioned, for both types of probe, three ports enable the electrical paths between the Vdd terminal of the circuit (port 1), the spying probe (port 2), and the DC pad (port 3).

\begin{figure}[t]
    \centering
    \includegraphics[width=.6\columnwidth]{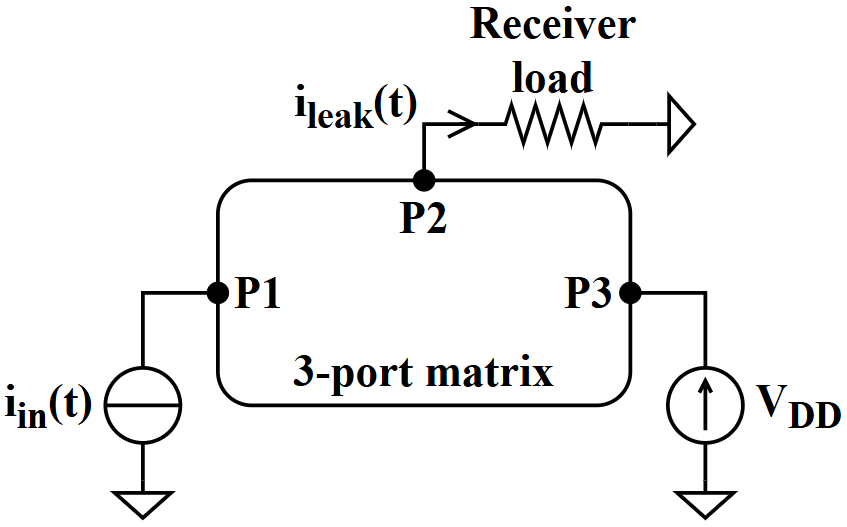}
    \caption{Electrical schematic for transient analysis of the attack.}
    \label{fig:Schematic_transient_analysis}
\end{figure}

The EM solver returns a three port matrix which models the coupling coefficients between ports from 0 to 200~GHz. 
A transient analysis is then performed following the schematic scenario presented in Fig.~\ref{fig:Schematic_transient_analysis}. The current source $i_{in}(t)$ connected to port 1 represents the current consumption of the victim circuit (reference trace mentioned in \ref{subsec:baseline_external_attack}). 
The voltage source "Vdd" at port 3 represents the DC power supply, although it has no effect on the results since it can be seen as a short-circuit for both RF signal and any of its harmonics.
The resistor labeled "receiver load" at port 2 represents the measurement system integrated in the observing chiplet (see Fig.~\ref{fig:system}). Its value is arbitrarily set to 50~ohm. The current $i_{leak}(t)$ flowing into the receiver load is the leakage current captured by the observing chiplet.

\begin{figure}[t]
    \centering
    \includegraphics[width=.8\columnwidth]{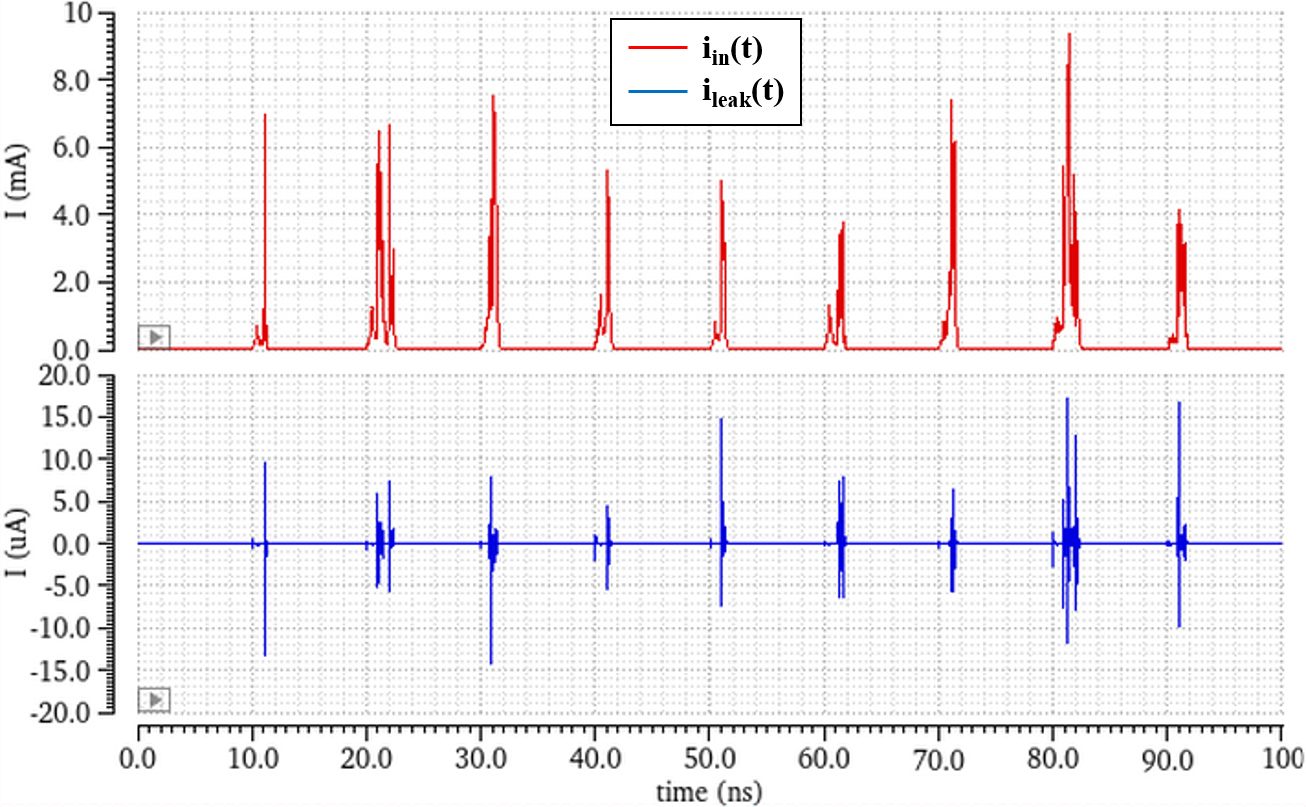}
    \caption{Transient analysis results using the capacitive probe matrix. Chiplets are spaced apart with 20~µm of air gap.\vspace{-0.4cm}}
    \label{fig:Transient_analysis_capacitive_probe}
\end{figure}

\begin{figure}[t]
    \centering
    \includegraphics[width=.8\columnwidth]{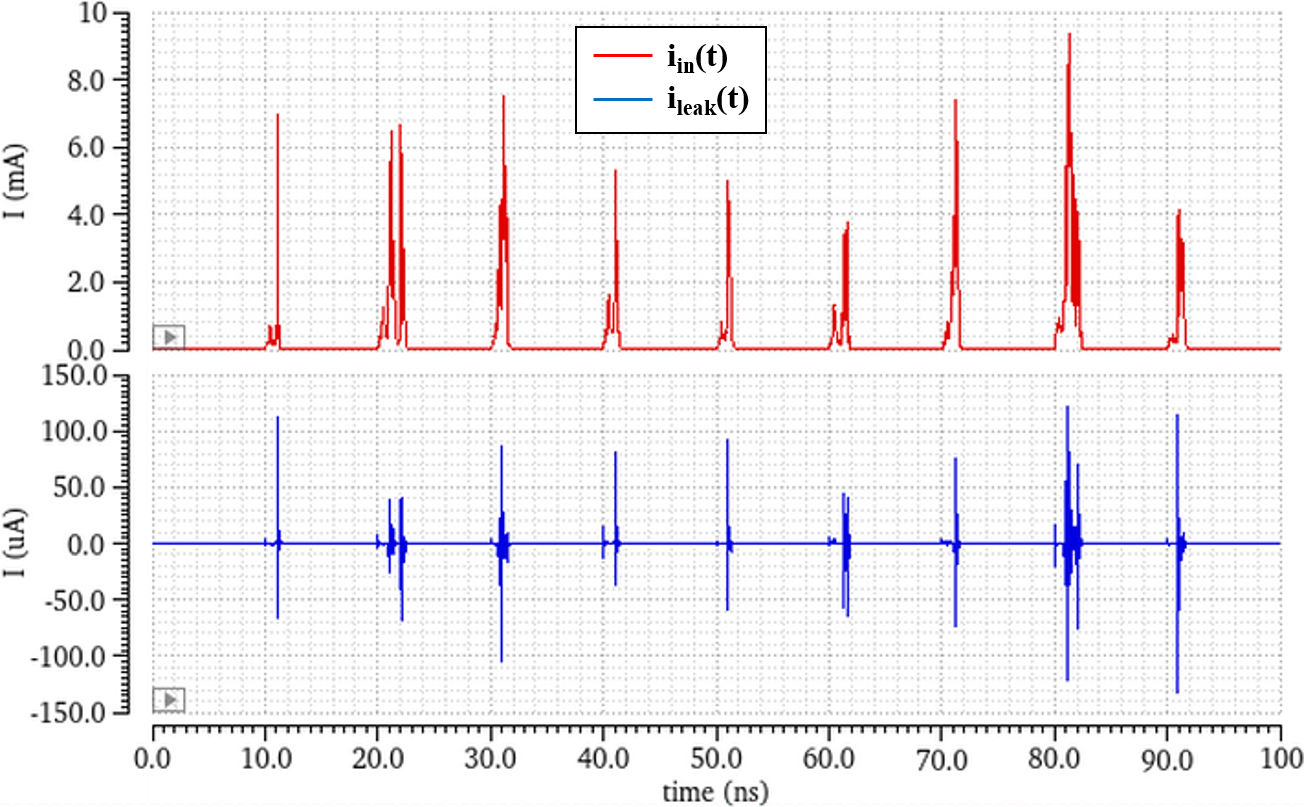}
    \caption{Transient analysis results using the inductive+capacitive probe matrix. Chiplets are spaced apart with 20~µm of air gap.}
    \label{fig:Transient_analysis_inductive_probe}
\end{figure}

The simulated results for $i_{in}(t)$ and $i_{leak}(t)$ are presented in Fig.~\ref{fig:Transient_analysis_capacitive_probe} and Fig.~\ref{fig:Transient_analysis_inductive_probe} for both capacitive and inductive+capacitive coupling, respectively. Both cases show that a leakage current is effectively coupled to the receiver load. The shape of the leakage trace is altered in comparison to the reference trace due to the filtering effect created by the spying probes.

\begin{figure}[t]
    \centering
    \includegraphics[width=\columnwidth]{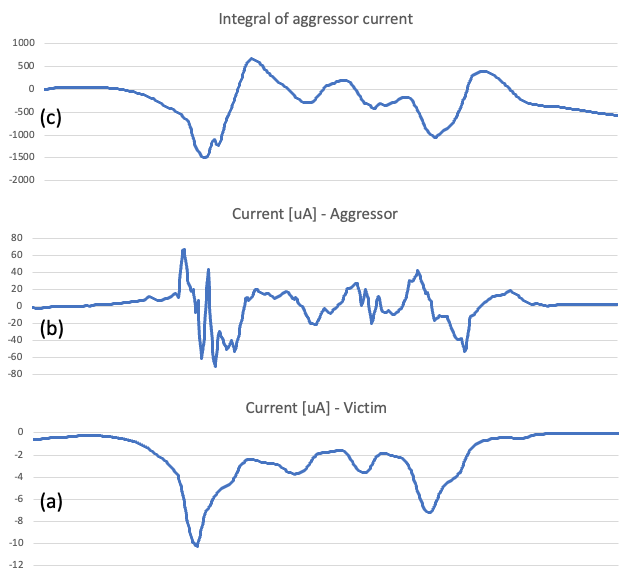}
    \caption{Waveforms associated with the capacitive coupling scenario. (a): the reference supply current measured on the victim chiplet. (b): the signal obtained on the observing chiplet after capacitive coupling, whose shape is close to the derivative of the original trace. (c): the integrated waveform used to recover a signal compatible with the subsequent side-channel analysis.}
    \label{fig:cap_coupling_waveforms}
\end{figure}

\subsection{Analysis of the Coupled and Digitized Signal}
\label{subsec:coupled_signal_analysis}

We now analyze the signal retrieved on the observing chiplet after cross-chiplet propagation and coupling. The objective is to determine whether the secret-dependent information remains exploitable once the leakage is no longer observed directly from the victim supply current, but instead through the signal coupled to the second chiplet. 

\subsubsection{Capacitive Coupling}
For the capacitive coupling case, the signal observed on the second chiplet does not directly reproduce the original supply current of the victim circuit. Instead, the coupling behaves as an RC high-pass effect, and the retrieved waveform is close to the temporal derivative of the original signal. As a result, directly applying the same DPA procedure as in Section~\ref{subsec:baseline_external_attack} to the raw coupled signal does not easily reveal the secret key.

This effect is illustrated in Fig.~\ref{fig:cap_coupling_waveforms}. The bottom plot (Fig.~\ref{fig:cap_coupling_waveforms}a) shows the reference supply current of the victim chiplet during one execution of the target operation. The middle plot (Fig.~\ref{fig:cap_coupling_waveforms}b) shows the signal obtained on the observing chiplet after capacitive coupling. It can be seen that this waveform closely resembles the derivative of the original current trace. In order to recover a signal more suitable for side-channel analysis, we therefore reconstruct an approximation of the original waveform by numerically integrating the coupled signal (Fig.~\ref{fig:cap_coupling_waveforms}c).

In practice, this reconstruction is intentionally kept simple. Let \(y[n]\) denote the discrete-time samples of the signal obtained after capacitive coupling. The reconstructed signal \(\hat{x}[n]\) is computed through cumulative summation:
\[
\hat{x}[n] = \hat{x}[n-1] + y[n],
\]
with \(\hat{x}[0]\) initialized to 0. Equivalently, each sample is obtained by adding the current sample to the accumulated value of the previous samples. The top plot of Fig.~\ref{fig:cap_coupling_waveforms} shows the resulting integrated waveform.

\begin{figure}[t]
    \centering
    \includegraphics[width=\columnwidth]{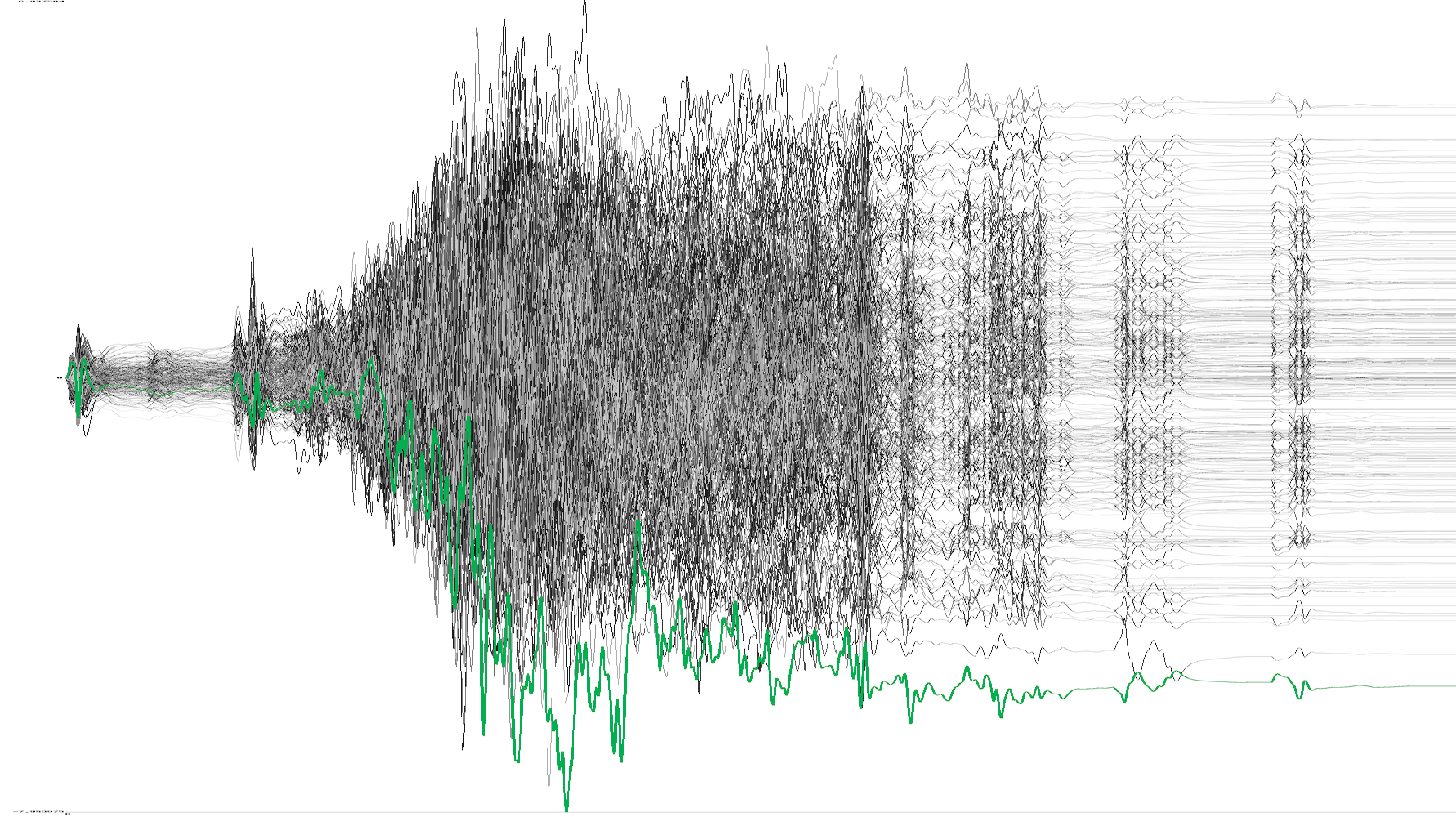}
    \caption{Outcome of the DPA applied to the reconstructed traces obtained in the capacitive coupling case.}
    \label{fig:cap_dpa_result}
\end{figure}

The same DPA procedure as in Section~\ref{subsec:baseline_external_attack} is then applied to the reconstructed traces, using the same target intermediate value and the same leakage model. The corresponding results are shown in Fig.~\ref{fig:cap_dpa_result}. The correct key hypothesis again exhibits the dominant distinguisher peak, showing that the key can still be recovered from the signal captured through the observing chiplet. This result demonstrates the feasibility of the proposed attack in the capacitive coupling scenario. The DPA curves are not perfectly aligned at the end of the simulation window. This behavior is expected. Indeed, the integral of a signal is defined up to an arbitrary additive constant, and different choices of this constant shift the reconstructed traces vertically without changing their validity. In principle, this offset could be adjusted to compensate for electrical noise and improve trace alignment, but such compensation was not applied here. Consequently, the reconstructed traces corresponding to different executions do not share exactly the same vertical offset.

\subsubsection{Inductive+Capacitive Coupling}
\label{subsubsec:inductive_coupling_results}

\begin{figure}[t]
    \centering
    \includegraphics[width=\columnwidth]{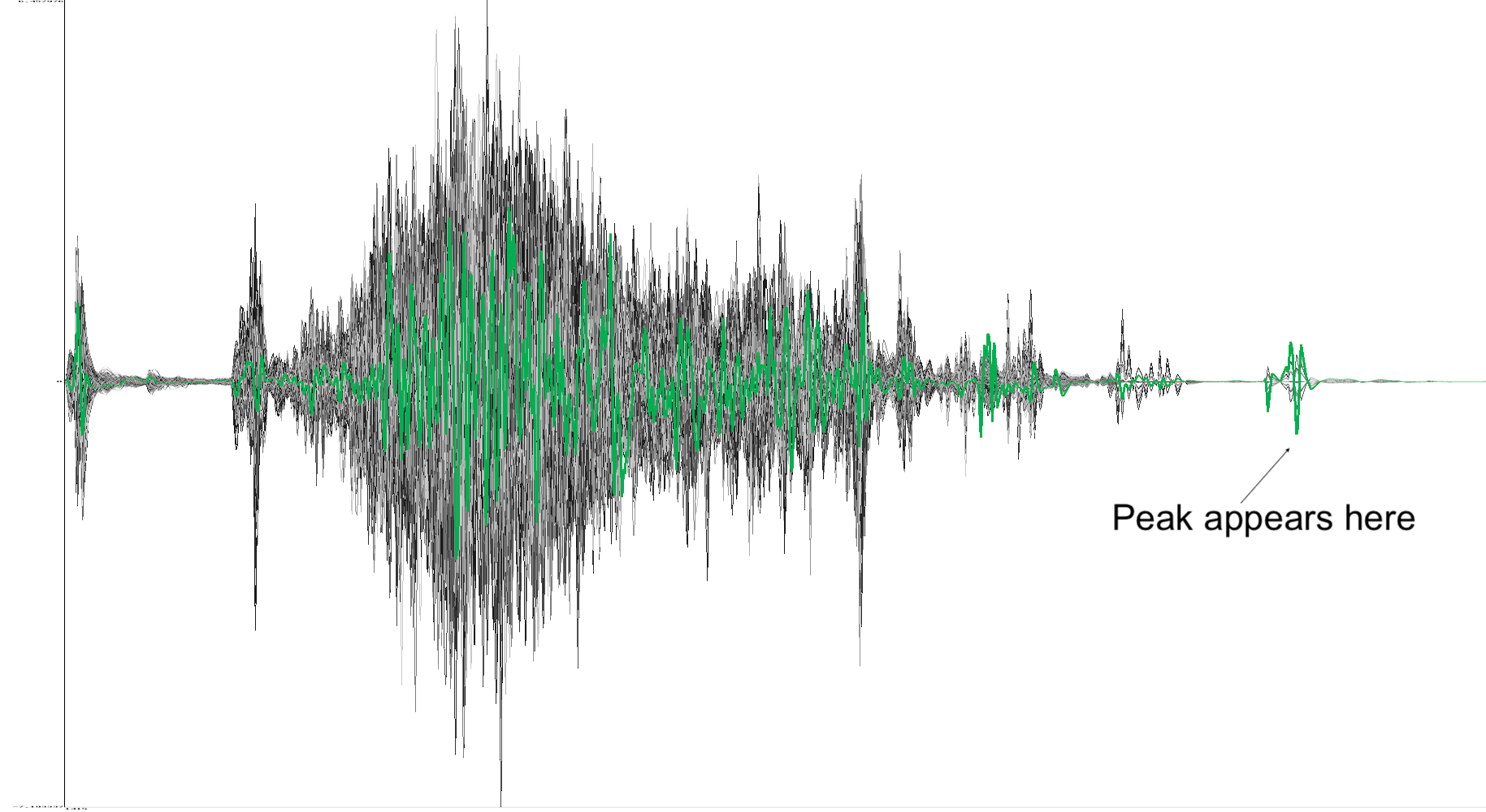}
    \caption{Outcome of the DPA applied to the reconstructed traces obtained in the inductive+capacitive coupling case.\vspace{-0.5cm}}
    \label{fig:ind_dpa_result}
\end{figure}

For the combined inductive and capacitive coupling case, no compensation or reconstruction step is applied. Unlike the purely capacitive case, for which the observed waveform can be approximated as the derivative of the original signal, the combined coupling mechanism produces a more complex transformation of the leakage. As a result, a simple correction is not straightforward. In this work, we therefore directly analyze the retrieved signal using the same DPA methodology as in the baseline experiment, while leaving the investigation of more accurate reconstruction or adaptation strategies for future work. The corresponding result is shown in Fig.~\ref{fig:ind_dpa_result}.

The results indicate that the secret key can still be recovered, but only when focusing on the very final part of the measurement, that is, when the actual transition of the S-box output takes place. In the earlier part of the waveform, the combined effects of the circuit activity, the propagation path, and the combined inductive and capacitive coupling mechanisms produce a more complex response in which the useful leakage is mixed with other components. As a consequence, the exploitable information is less directly visible during the first part of the trace and becomes clearly distinguishable only near the final switching event associated with the sensitive operation.

These observations highlight an important point: although the proposed cross-chiplet attack remains feasible, the shape of the retrieved leakage strongly depends on the physical coupling mechanism. In the capacitive case, signal reconstruction through integration is required to recover a waveform compatible with the original leakage model. In the more complex case, the raw coupled signal can still be attacked, but the exploitable leakage is concentrated in a narrower temporal region. Therefore, the attack methodology must be adapted to the physical effect exploited by the communication or coupling structure.

\section{Limitations}
\label{sec:limitations}

To conclude on the EM coupling study, several hypotheses should be discussed. 
First, the custom material stack (Fig.~\ref{fig:stack_custom}) does not correspond to an actual IC technology Back-End-Of-Line (BEOL). Various simplifications are made on the thicknesses and electric properties (e.g., resistivity or permittivity) of the materials and could not reflect well an actual BEOL. The assumptions on the medium characteristics between the two chiplets are also simplified. 

Second, the geometries of the power supply line and ground plane are greatly simplified (Fig.~\ref{fig:Capacitive_probe_3Dview}). A more in-depth study should reuse an actual power supply routing, including the DC pads and wire-bondings. Moreover, an actual power supply routing might integrate decoupling capacitors between the supply line and the ground plane precisely to avoid signal leakage and variations of the Vdd. The shape and position of the spying probes are also simplified and do not correspond to a realistic scenario as imagined in \ref{subsec:Communication-Oriented_Chiplets}. 

Lastly, the schematic of Fig.~\ref{fig:Schematic_transient_analysis} does not consider the actual frequency dependent impedances presented to the 3-port matrix. The Vdd terminal of the victim circuit can be seen as a real signal source and should then include a parallel impedance. A wire-bond connected to the DC pad should present an inductive impedance, and the receiver load should either be matched to the spying probe impedance or present an high impedance input (typically the case for ADCs).

Therefore, the results presented in Fig.~\ref{fig:Transient_analysis_capacitive_probe} and Fig.~\ref{fig:Transient_analysis_inductive_probe} only prove that a signal leakage from the Vdd line is possible, and caution is advised about the power and integrity of the leaked signal in the receiver load.
%etre prudents sur : puissance couplée, effet de filtrage et intégrité du signal couplé

\section{Conclusion}
\label{sec:conclusion}
This work investigated whether side-channel information can be observed across chiplets in a heterogeneous integrated system. The study shows that secret-dependent information may remain accessible even after cross-chiplet propagation and acquisition on a neighboring chiplet, which confirms the practicality of the considered attack scenario. More broadly, the results suggest that communication-oriented chiplets may act not only as functional interfaces, but also as side-channel observation platforms.

An important outcome of this study is that the leakage observed after propagation cannot be analyzed as a simple replica of the original signal. Instead, the coupling mechanism may significantly transform the waveform, so that the leakage model must be adapted to the underlying physical effect. In the considered cases, this was illustrated by the need to account for the signal transformation induced by the coupling path before successfully recovering the secret. This observation is particularly relevant for future attacks, since it shows that cross-chiplet side-channel analysis is not only a matter of signal capture, but also of properly understanding and modeling the physical channel between the victim and the observer.

These findings open several research directions. On the offensive side, they motivate the development of more accurate signal models, reconstruction techniques, and attack methodologies specifically tailored to cross-chiplet leakage. On the defensive side, they highlight the need for new countermeasures targeting physical leakage paths created by advanced integration, including communication interfaces, package-level coupling effects, and the relative placement of heterogeneous chiplets. Overall, this paper points to a broader security implication of heterogeneous integration: in future chiplet-based systems, the boundary between communication structures and sensing structures may become blurred. Taking this dual-use risk into account will be essential for the secure design of next-generation multi-chiplet architectures.

\balance
\bibliographystyle{IEEEtran}
\bibliography{refs}
\end{document}